# Raman Spectroscopy for the Early Detection of Huanglongbing and Citrus Canker in Plants: A Review

[1]Ashis Kumar Das, [2]Shikha Sharma, [3]Snehprabha Gujarathi, [4]Sushmita Mena, [5]Saurav Bharadwaj

[1]ICAR-Central Citrus Research Institute, Maharashtra, India

[2]CSK Himachal Pradesh Agricultural University, Himachal Pradesh, India

[3,5]Parul Institute of Engineering and Technology, Parul University, Gujarat, India

[4]Parul Institute of Applied Sciences, Parul University, Gujarat, India

[5]Corresponding Author: Saurav Bharadwaj, saurav.bharadwaj33162@paruluniversity.ac.in,

[5]ORCID: 0000-0003-1882-7287

**Abstract:** Citrus crops are of immense economic and agricultural importance worldwide, but are highly vulnerable to destructive diseases such as Huanglongbing (citrus greening) and citrus canker. These diseases often remain asymptomatic in early stages, making timely diagnosis difficult and resulting in substantial yield and quality losses. Rapid, non-invasive, and accurate early detection methods are therefore essential for effective disease management and sustainable citrus production. Raman spectroscopy, with its molecular specificity, non-destructive nature, and minimal sample preparation requirements, has emerged as a promising diagnostic technology for this purpose. This review outlines recent advances in applying Raman spectroscopy for the early detection of Huanglongbing and citrus canker, with a focus on spectral biomarkers that reflect pathogen-induced physiological and biochemical alterations in citrus tissues. Comparative insights are provided between portable, in-field Raman devices and conventional laboratory-based systems, highlighting diagnostic accuracy, operational feasibility, and deployment potential. The integration of chemometric and machine learning techniques for enhanced classification and automated disease recognition is examined. By consolidating current research, this review underscores the potential of Raman spectroscopy as a field-deployable solution for precision agriculture. It discusses future challenges and opportunities for the large-scale adoption of this technology in citrus orchard monitoring.

# 1 Introduction

Citrus diseases represent a significant threat to global citrus production, with profound impacts on both yield and the economic viability of citrus crops such as oranges, lemons, limes, and grapefruits [1]. These diseases arise from a wide spectrum of pathogens, including bacteria, fungi, viruses, and nematodes, as well as abiotic stressors like nutrient imbalance and drought [2]. Among them, Huanglongbing, or citrus greening, is the most destructive [3]. Huanglongbing is caused by phloem-limited bacteria belonging to the genus *Candidatus Liberibacter*, including *Candidatus Liberibacter asiaticus* (CLas), *Candidatus Liberibacter africanus* (CLaf), and *Candidatus Liberibacter americanus* (CLam). The disease is vectored by two main insect carriers: the Asian citrus psyllid (*Diaphorina citri*) and the African citrus psyllid (*Trioza erytreae*) [4], [5]. Infected trees exhibit characteristic symptoms such as blotchy mottled leaves, yellowing shoots, lopsided and bitter fruits, and progressive decline leading to eventual tree death [6]. Given the lack of a definitive cure for Huanglongbing, management strategies rely on early diagnosis, removal of infected trees, and strict vector control using insecticides and biological agents [7], [8], [9]. Another prominent bacterial disease affecting citrus is Citrus Canker, caused by *Xanthomonas citri*. This pathogen induces necrotic lesions on leaves, stems, and fruit, which are often surrounded by a distinctive yellow halo [10], [11].

Effective control of citrus diseases necessitates the deployment of integrated disease management (IDM) practices that address both pathogen and vector control while promoting plant resilience [12], [13]. The use of disease-resistant rootstocks and scions, achieved through conventional breeding or molecular genetic techniques, provides a sustainable long-term solution [14]. Cultural practices such as site selection, optimized irrigation, balanced fertilization, sanitation of pruning tools, and removal of symptomatic trees reduce pathogen load and promote healthy canopy structure [15]. Regulatory frameworks also play a critical role; quarantine enforcement, phytosanitary inspections, and movement restrictions are essential for preventing pathogen introduction into disease-free zones [16], [17]. Chemical control remains a widely used tactic, with copper-based formulations proving effective against citrus canker and insecticides targeting psyllid vectors of Huanglongbing [8], [18]. Overreliance on agrochemicals poses risks such as pesticide resistance and ecological harm [19]. Judicious use of chemicals within an Integrated Pest Management (IPM) framework is essential [15]. Biological control strategies are also gaining prominence, employing beneficial microbes to colonize the rhizosphere or phyllosphere, and natural enemies like lady beetles, lacewings, and parasitic wasps to suppress pest populations [5], [8], [20].

Despite being the most immediate method for field-level disease assessment, visual diagnosis of citrus diseases is inherently limited by the nonspecific and overlapping nature of symptoms [21]. Multiple pathogens produce similar visual cues—such as chlorosis, necrosis, and leaf curling—making it difficult to accurately attribute symptoms to specific diseases [22], [23]. Environmental variables such as lighting, leaf age, cultivar-dependent morphology, and canopy shading further obscure symptom clarity in field imagery [24]. Early-stage infections, often asymptomatic, evade detection altogether, while the variability in symptom presentation compromises the reliability of automated image-based detection models [25], [26].

Molecular diagnostic techniques have emerged as essential tools for accurate and early detection of citrus pathogens, enabling intervention before significant crop damage occurs [27], [28]. These techniques, primarily targeting pathogen-specific nucleic acid sequences, offer high sensitivity and specificity. Polymerase Chain Reaction (PCR) is the cornerstone of molecular diagnostics, facilitating the amplification of DNA fragments unique to specific pathogens [1], [29]. For RNA viruses, Reverse Transcription PCR (RT-PCR) is used to synthesize cDNA, which is then amplified. Advanced variants such as nested PCR and multiplex PCR provide increased sensitivity and allow for simultaneous detection of multiple pathogens, respectively [30], [31]. While older techniques like Southern blotting and dot blot hybridization once served in diagnostic workflows, they have largely been superseded by PCR-based methods due to their greater efficiency [32], [33]. Restriction Fragment Length Polymorphism (RFLP) remains useful for differentiating between pathogen strains based on DNA polymorphisms [34]. A promising newer approach is Loop-mediated Isothermal Amplification (LAMP), which offers rapid and field-deployable diagnostics by enabling DNA amplification at a constant temperature without the need for thermocyclers [35], [36], [37].

In parallel with molecular methods, optical diagnostic technologies are gaining traction as non-invasive, rapid, and cost-effective solutions for early detection of citrus diseases [7], [38]. Techniques like Raman spectroscopy detect biochemical changes in plant tissues induced by pathogen infection, often before visible symptoms manifest. Raman spectroscopy offers high spectral resolution, minimal sample preparation, and in situ measurement capabilities, making it suitable for real-time disease surveillance across large orchards [39], [40]. However, the application of optical methods in real-world agricultural settings remains limited. Environmental factors such as light variability, leaf surface properties, and spectral noise introduce challenges in signal interpretation [41], [42]. Overcoming these barriers requires the development of portable, field-optimized devices integrated with machine learning algorithms for real-time data processing [43], [44]. Advances in miniaturized Raman systems and the coupling of spectral data with precision agriculture platforms can significantly enhance decision-making in orchard management. The development of standardized protocols and validation against diverse field conditions is essential to ensure the reliability and scalability of optical diagnostics [7], [45].

This review provides a critical assessment of the current status and practical potential of Raman spectroscopy as a non-destructive, label-free diagnostic tool for the early detection of citrus diseases under real-world orchard conditions. While extensive research has demonstrated the utility of Raman-based methods in detecting pathogen-induced biochemical changes at the leaf or fruit level, much of this work has been limited to controlled laboratory settings, with relatively few studies addressing large-scale field deployment. Recent technological advances have improved the sensitivity and specificity of Raman systems, allowing for the identification of pathogen-specific biomolecular markers relevant to plant health. When compared to conventional diagnostics and other optical techniques, Raman spectroscopy offers superior chemical resolution and faster detection times. However, its practical application is constrained by several factors, including signal noise from environmental conditions, physiological heterogeneity among plants, and the lack of low-cost, portable devices optimized for field use. Critical gaps remain in the integration of Raman-derived data into agronomic decision-making frameworks, as well as in the development of standardized, reproducible field protocols.

## 2 Citrus Diseases

### 2.1 Huanglongbing (Citrus Greening)

Huanglongbing is a devastating citrus disease primarily caused by the phloem-limited bacterium *Candidatus Liberibacter* species, which is transmitted by psyllid insects [46], [47]. The disease cycle begins when an infected psyllid feeds on the phloem of a healthy citrus plant, introducing the bacterium into the vascular system [48]. Inside the host, the bacterium colonizes the phloem tissue, disrupting nutrient transport and inducing systemic symptoms such as leaf mottling, stunted growth, premature fruit drop, and deformed, bitter fruit [22], [49]. Despite its eventual severe impact, the bacterium spreads slowly and unevenly within the plant, often leading to a prolonged asymptomatic phase [50]. In the insect vector, the bacterium multiplies in the gut and salivary glands, enabling persistent and propagative transmission [51], [52]. Once acquired, psyllids remain infective for life. Nymphs are particularly efficient at acquiring the bacterium, which undergoes a latent period within the insect before it becomes transmissible, as shown in Figure 1 [53], [54].

Long-distance spread mainly occurs through the movement of infected plant material or psyllids, while local dissemination is driven by psyllid flight or wind-assisted dispersal [26], [42]. The disease can also be transmitted through grafting with infected budwood. Management of Huanglongbing is extremely challenging due to the systemic nature of the pathogen, the cryptic behavior of the vector, and the absence of effective treatments [39]. The disease cycle is sustained by infected trees that act as bacterial reservoirs and by the psyllid capacity for survival and dispersal, facilitating rapid spread within and between orchards [55]. Cultural practices such as the removal of infected trees, vector control through insecticide application, use of certified disease-free nursery stock, and systemic treatments can reduce disease incidence, but none are curative. Huanglongbing remains a significant obstacle to integrated citrus disease management strategies [56].

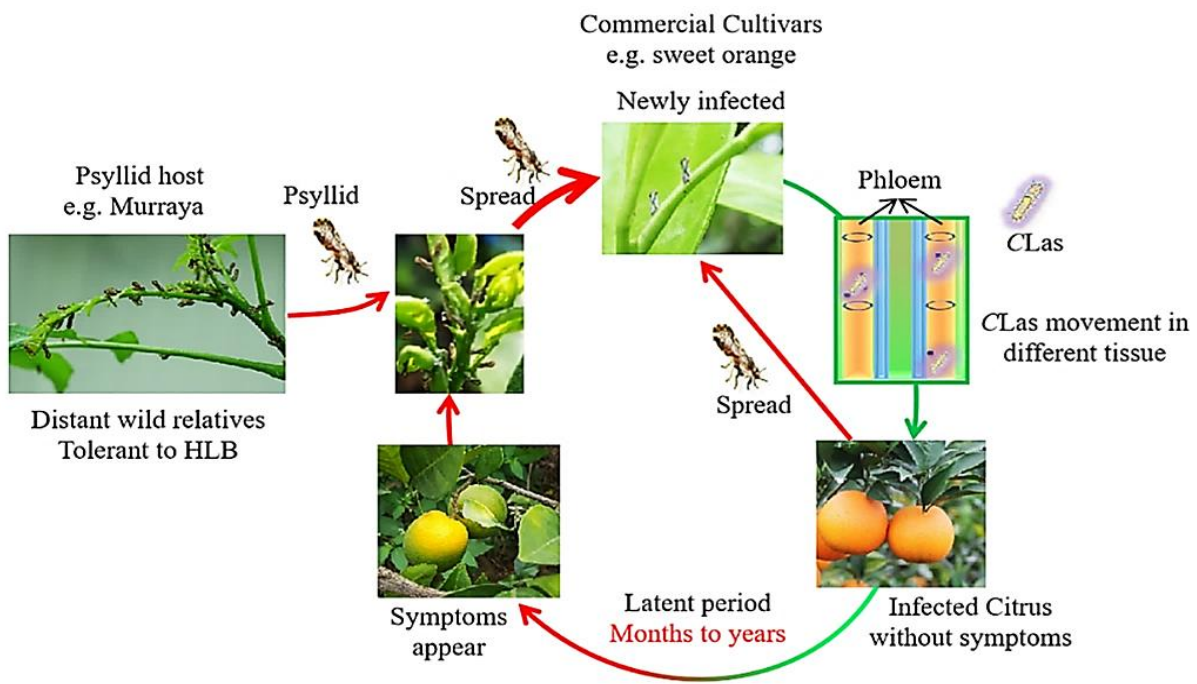

**Figure 1:** Disease cycle of Huanglongbing involving transmission by psyllid vectors from tolerant wild hosts to commercial citrus cultivars. The bacterium spreads through the phloem, causing latent infections followed by symptom development and further spread [51].

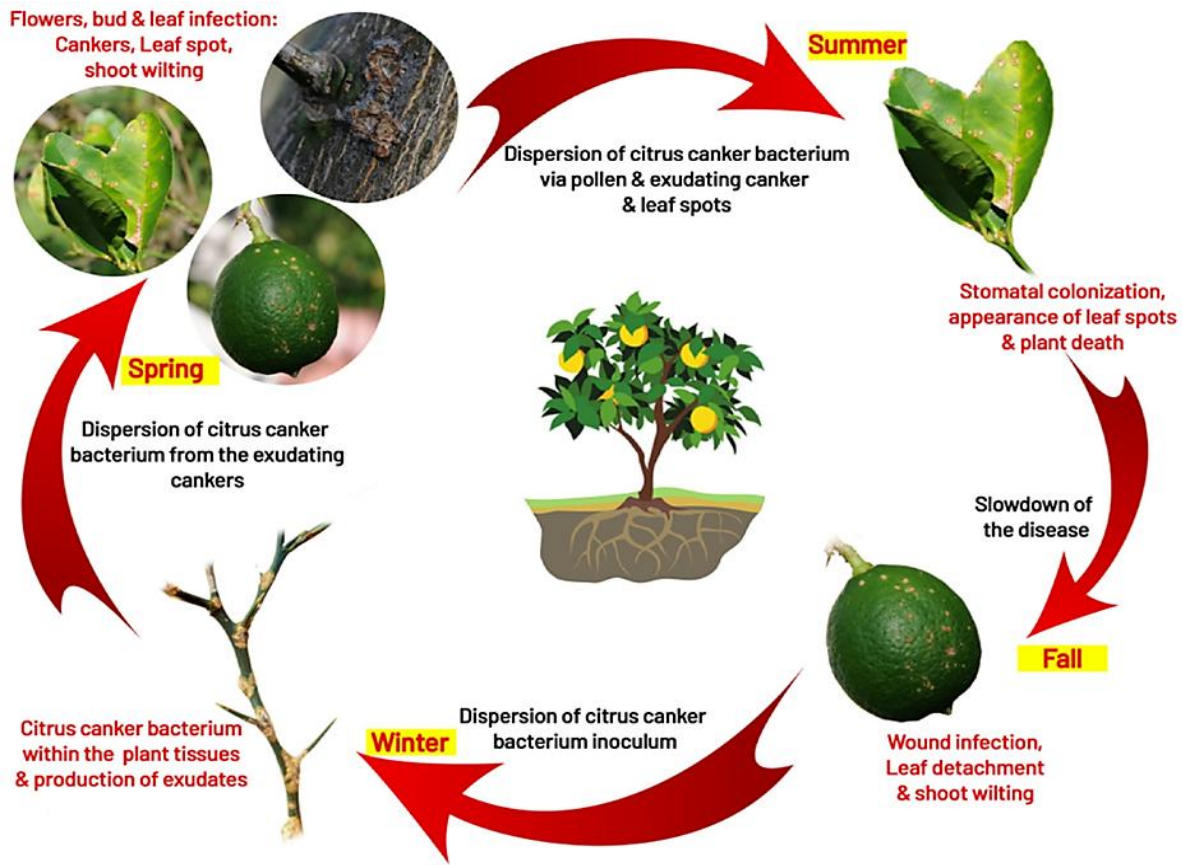

**Figure 2:** Seasonal disease cycle of citrus canker, illustrating bacterial dispersion, infection, and symptom development across spring, summer, fall, and winter. The bacterium spreads via exudates, pollen, and lesions, leading to cankers, leaf drop, shoot dieback, and overall plant decline, with peak disease activity during warmer seasons [10].

## 2.2 Citrus Canker

Citrus canker is a polycyclic disease characterized by multiple infection cycles within a single growing season. The causal bacterium (*Xanthomonas citri*) survives epiphytically on leaf surfaces and in infected plant tissues, such as lesions on twigs, stems, and fruit [11]. The primary inoculum is dispersed mainly via wind-driven rain, which facilitates bacterial entry into host tissues through natural openings (e.g., stomata) or wounds caused by wind, insect activity, or mechanical injury, as shown in Figure 2 [10]. Once inside the host, the bacterium colonizes the intercellular spaces, leading to the development of necrotic, raised lesions with water-soaked margins and a yellow halo—particularly on young, actively growing tissues. Under favorable conditions—warm temperatures (20–30°C) and high humidity—lesions exude bacterial ooze, which acts as a source of secondary inoculum [11].

Disease spread accelerates during periods of frequent rainfall and high winds, especially during the citrus flushing period. Infected tissues serve as reservoirs for the pathogen, allowing it to persist and reemerge in subsequent seasons [10], [57]. Effective management of citrus canker requires an integrated approach, including the use of resistant cultivars, the establishment of windbreaks, application of copper-based bactericides, and implementation of strict phytosanitary measures [10].

**3 Raman Instrumentation**

**3.1 Raman Spectroscopy**

Raman spectroscopy is an analytical technique based on the inelastic scattering of monochromatic light—typically from a laser source—that provides molecular fingerprint information about a sample [58]. The core instrumentation of a Raman spectroscope includes a laser excitation source, sample illumination optics, collection optics, a wavelength-dispersive element, and a detector system, as shown in Figure 3 [41]. The laser source is generally a continuous-wave laser operating in the visible, near-infrared (NIR), or ultraviolet (UV) spectral range, chosen to optimize Raman signal intensity while minimizing fluorescence interference from the sample [59]. Common laser wavelengths include 532 nm, 633 nm, and 785 nm, with the selection depending on the sample properties and the desired spectral resolution [60]. The excitation beam is directed onto the sample through a microscope objective or focusing lens, ensuring efficient delivery of laser power to a small focal volume, thereby enhancing the Raman signal [61]. The scattered light—comprising predominantly Rayleigh (elastic) and Raman (inelastic) components—is typically collected in a backscattering geometry to maximize signal collection efficiency. A set of optical filters, such as notch or edge filters, is employed to suppress the intense Rayleigh scattered light while transmitting the weaker Raman-shifted photons. This filtering is critical for achieving a high signal-to-noise ratio in the acquired spectra [62]. The filtered Raman-scattered light is then fed into a spectrometer, which disperses the light by wavelength. The spectrometer generally consists of a diffraction grating coupled with a focal plane array detector, such as a charge-coupled device (CCD) [63]. The groove density and blaze wavelength of the diffraction grating are selected to optimize spectral resolution and throughput within the Raman spectral range of interest [64]. The CCD detector is cooled to reduce thermal noise and converts the dispersed photons into electrical signals, which are subsequently digitized and processed by specialized software to generate Raman spectra [65].

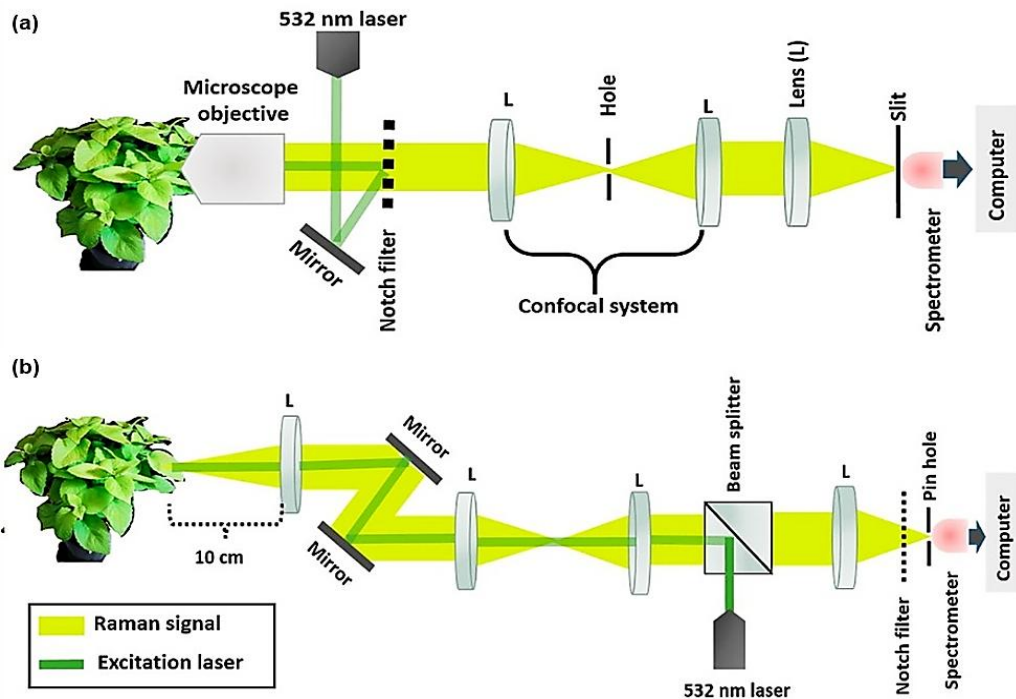

**Figure 3:** Experimental set-up of (a) a confocal Raman microscopic system using a 532 nm laser for localized analysis and (b) a remote Raman spectroscopic system enabling standoff detection of plant signals at a 10 cm distance [41].

## 3.2 Raman Imaging

Raman imaging extends conventional Raman spectroscopy by enabling spatially resolved chemical mapping of samples, thereby providing both spectral and morphological information simultaneously [64], [66]. The instrumentation for Raman imaging builds upon the basic Raman spectroscopy setup by incorporating precise scanning mechanisms and advanced data acquisition systems [67]. A confocal Raman microscope configuration is typically used, in which a high numerical aperture (NA) objective focuses the excitation laser to a diffraction-limited spot on the sample, as shown in Figure 4 [68], [69]. The sample is scanned either by deflecting the laser beam using galvanometric mirrors or by moving the sample stage with high-precision motorized actuators. This allows for systematic collection of Raman spectra across a defined grid or area [70]. At each scanned position, the collected Raman-scattered light is spectrally dispersed and detected, producing a hyperspectral dataset—a three-dimensional data cube containing spatial coordinates and corresponding Raman spectra [71]. This data cube enables the reconstruction of two-dimensional chemical images by extracting intensity maps of specific Raman bands that are characteristic of molecular species or functional groups [62]. The choice of laser wavelength affects both the Raman scattering efficiency and the achievable spatial resolution. Shorter wavelengths generally provide higher spatial resolution due to smaller diffraction-limited spot sizes, but they can also increase fluorescence background and the risk of photodamage [70], [72]. The objective lens plays a crucial role in determining the NA, which directly influences both the focusing ability of the excitation beam and the efficiency of scattered light collection [73]. Higher NA objectives allow for tighter focusing, improving spatial resolution and signal intensity, though at the cost of reduced depth of focus. The spectrometer design—particularly the groove density of the diffraction grating and the overall optical layout—governs spectral dispersion and, consequently, spectral resolution, which is the ability to distinguish closely spaced Raman peaks [64]. Higher spectral resolution improves the accuracy of molecular vibration identification but may necessitate longer integration times, affecting scanning speed [74]. Detector sensitivity and noise characteristics—typically associated with cooled CCD or complementary metal-oxide-semiconductor (CMOS) detectors—determine how rapidly and accurately weak Raman signals can be recorded, thereby impacting both data quality and acquisition speed [75].

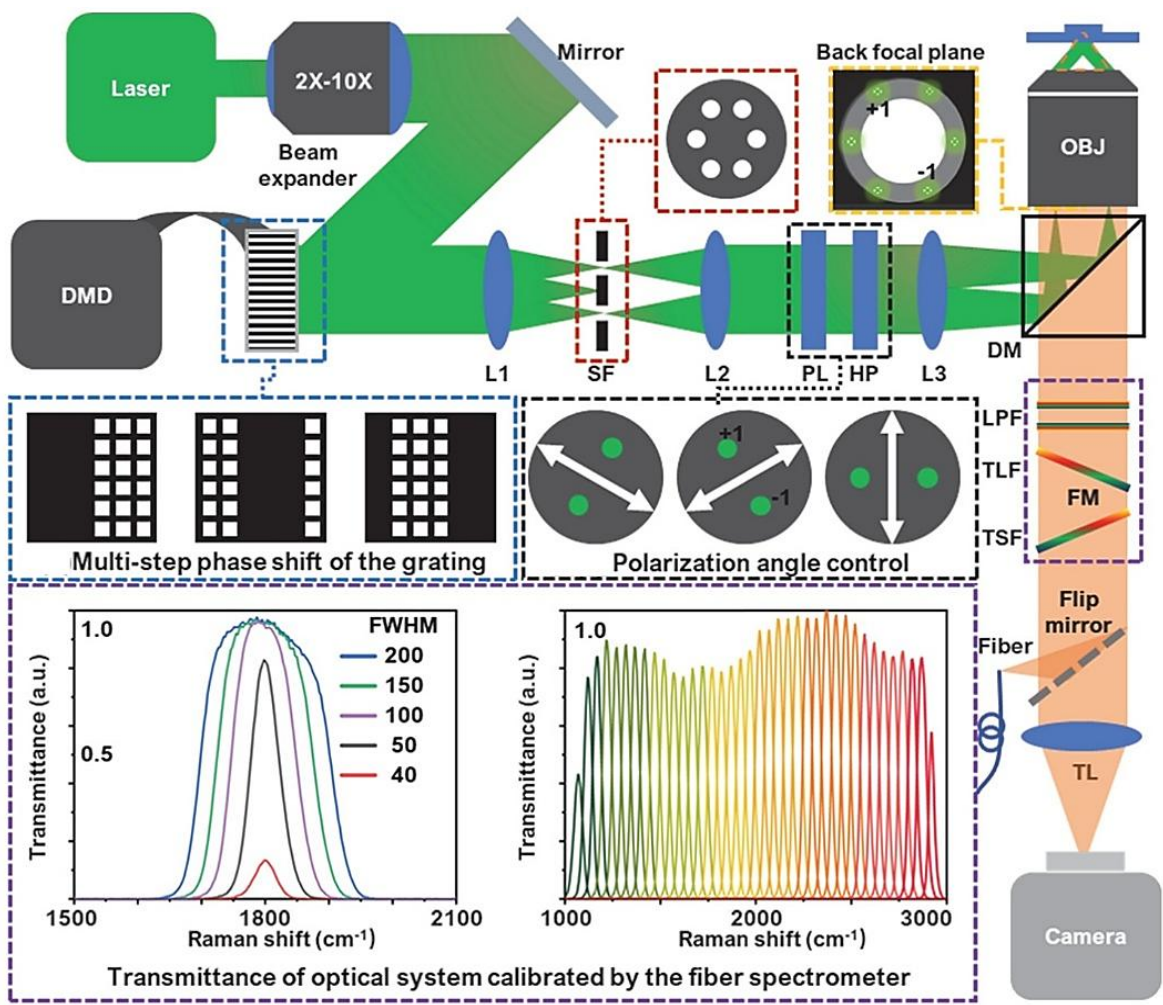

**Figure 4:** Schematic diagram of the structured illumination Raman microscopy system [68].

**3.3 Pre-processing Techniques**

The preprocessing is a critical step in Raman spectroscopy and Raman imaging, significantly improving the quality and interpretability of spectral data. Raw Raman signals are often affected by various artefacts, including cosmic rays, fluorescence background, baseline drift, and instrumental noise [76]. These distortions can obscure meaningful vibrational features and compromise downstream analyses such as multivariate statistical modeling, clustering, and machine learning classification [77]. The preprocessing transforms raw data into a standardized format suitable for both qualitative and quantitative interpretation [78].

Baseline correction is one of the most essential preprocessing steps, aiming to eliminate background fluorescence and baseline drift that can mask genuine Raman peaks [79]. Several algorithms are commonly used for this purpose. Polynomial fitting involves fitting a low-order polynomial to the baseline and subtracting it. Asymmetric Least Squares iteratively balances signal fidelity with baseline smoothness. Adaptive Iteratively Reweighted Penalized Least Squares further refines baseline estimation by adaptively weighting spectral points during fitting. These methods effectively isolate true vibrational features by minimizing non-Raman background contributions. Smoothing techniques are employed to reduce high-frequency noise while preserving critical peak features—particularly important for low-intensity Raman signals or hyperspectral Raman images [80]. The Savitzky-Golay filter performs local polynomial regression within a moving window, smoothing the spectrum while maintaining peak shapes [81]. A simpler alternative, the moving average filter, averages signal intensity over a defined window. By mitigating noise, these methods enhance peak detectability without distorting the spectral structure [82].

Normalization corrects for signal intensity variations due to differences in sample concentration, laser power, acquisition time, or optical alignment. It ensures that spectra from different samples or spatial locations are comparable on a consistent scale [83]. Common normalization methods include total area normalization (dividing each spectrum by its total integrated area), vector normalization (scaling each spectrum to unit vector length), and Standard Normal Variate (SNV) transformation (centering and scaling each spectrum by its mean and standard deviation). These techniques are particularly valuable for comparative studies and chemometric analysis across large datasets [84].

Cosmic ray removal is essential, especially in Raman imaging, where thousands of spectra are acquired. Cosmic rays introduce sharp, non-reproducible spikes that can distort spectral interpretation. Median filtering replaces outlier points by comparing each data point with its neighbors [85]. Statistical outlier detection methods flag and correct points with abnormally high intensities. Effective cosmic ray removal is critical for ensuring data integrity and robust downstream analysis. Spatial filtering and denoising are particularly important in Raman imaging, where each pixel represents an individual Raman spectrum [86]. Image-level preprocessing improves visualization and reduces spatial noise. Spatial filters such as Gaussian and median filters are used to smooth image textures and minimize pixel-level artefacts. Multivariate denoising techniques distinguish signal from noise based on spectral variance across the image. These methods enable more accurate spatial interpretation of chemical distributions [64], [87].

## 4 Practical Applications

Raman spectroscopy is emerging as a powerful tool in precision agriculture for rapid, non-destructive diagnosis of plant stress [88]. In a study conducted by Sanchez et al., Raman spectroscopy was evaluated for its potential in early detection of Huanglongbing in citrus trees. Leaf samples were collected from healthy greenhouse-grown trees, in-field healthy trees, nutrient-deficient trees, and asymptomatic Huanglongbing -infected trees. Spectral data were obtained using an 830 nm laser operating at 495 mW with a 1-second acquisition time. Baseline correction was applied to the spectra before analysis. Raman spectroscopy revealed unique spectral features in in-field healthy trees that tested negative by qPCR, suggesting the presence of early-stage Huanglongbing that eluded detection by conventional molecular techniques. Significant spectral changes included increased intensity at 1610 cm$^{-1}$, attributed to lignin and phenylpropanoids, decreased intensity at 1525 cm$^{-1}$ linked to carotenoids, and shifts in $CH_2$/$CH_3$ vibrational bands. While nutrient-deficient and Huanglongbing -infected plants exhibited overlapping characteristics, the 1247 cm$^{-1}$ band was found exclusively in nutrient-deficient leaves. Multivariate statistical analysis supported these findings, demonstrating that Raman spectroscopy can not only detect Huanglongbing earlier than qPCR but also distinguish it from abiotic stress, such as nutrient deficiency, as illustrated in Figure 5 [44].

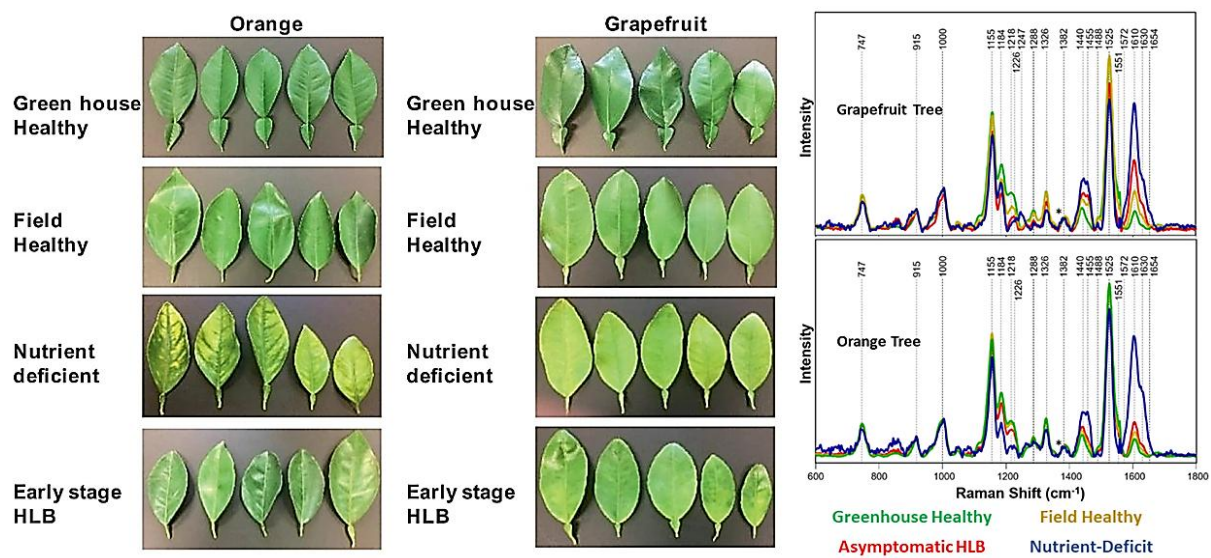

**Figure 5:** Representative leaves from healthy orange and grapefruit trees grown in greenhouse and field conditions, nutrient-deficient trees, and trees with early-stage Huanglongbing infection. Raman spectral variations highlight biochemical changes associated with nutrient deficiency and early asymptomatic Huanglongbing infection [44].

Biotic and abiotic stresses have profound effects on plant biochemistry, significantly altering the secondary metabolite profile. These biochemical shifts serve as vital stress indicators and can be precisely characterized using high-performance liquid chromatography (HPLC) and HPLC coupled with mass spectrometry (HPLC-MS). In the context of Huanglongbing infection, these techniques have revealed substantial changes in citrus leaf metabolite composition. HPLC-MS analysis demonstrated elevated levels of hydroxycinnamates, especially p-coumarates, and flavones such as vitexin-2-O-rhaminozide—compounds indicative of an activated phenylpropanoid pathway, a key defense mechanism. In asymptomatic, Huanglongbing -positive grapefruit leaves, enhanced vibrational bands at 1601 and 1630 cm$^{-1}$—associated with polyphenolic compounds like p-coumaric acid—were observed. Additionally, a shoulder at 1575 cm$^{-1}$ correlated with flavones such as vitexin-2-O-rhaminozide and diosmetin, known to accumulate in response to pathogenic stress. Concurrently, decreased Raman band intensities at 1000, 1155–1226, and 1525 cm$^{-1}$ suggested degradation of carotenoids like lutein, possibly due to oxidative stress or metabolic flux redirection. Spectral shifts from 1606 to 1601 cm$^{-1}$ while the 1630 cm$^{-1}$ band remained stable emphasize the biochemical specificity of Huanglongbing -induced stress. These findings confirm Raman spectroscopy, when integrated with HPLC-MS, as a robust tool for identifying stress-induced metabolic alterations in citrus, as shown in Figure 6 [89].

A portable field Raman spectroscopy system combined with multivariate statistical analysis was employed to evaluate its effectiveness in early Huanglongbing detection in Sweet Orange trees. Leaf samples, both symptomatic and asymptomatic, were collected from Mexican and Persian Lime trees across six municipalities—three in Colima and three in Jalisco, known Huanglongbing hotspots. Raman spectral analysis identified distinct biochemical anomalies in Huanglongbing -positive samples. Vibrational band differences were observed in carbohydrate-related regions (905, 1043, 1127, 1208, 1272, 1340, 1370, and 1260–1280 cm$^{-1}$), amino acid and protein regions (815, 830, 852, 918, 926, 970, 1002, 1053, and 1446 cm$^{-1}$), and lipid-associated regions (1734–1746 cm$^{-1}$). These spectral differences facilitated clear differentiation between healthy and Huanglongbing -infected plants. A Principal Component Analysis–Linear Discriminant Analysis (PCA–LDA) model was developed and demonstrated a high diagnostic performance, with a sensitivity of 86.9%, specificity of 91.4%, and precision of 89.2%. These outcomes validate the reliability of Raman spectroscopy as a field-deployable, non-invasive diagnostic tool for early Huanglongbing detection in citrus trees, offering significant potential for timely disease management and containment across affected regions [90].

Raman spectroscopy, being a nondestructive and noninvasive analytical method, allows for accurate identification of molecular structures in plant tissues [91]. In this study, a handheld Raman system, supported by chemometric models, effectively differentiated healthy citrus trees from those infected with Huanglongbing at early and late stages, as well as from nutrient-deficient trees. Diagnostic accuracy reached nearly 98% for grapefruit and 87% for orange trees. Classification accuracy for early vs late-stage Huanglongbing was particularly high—100% in grapefruit and about 94% in oranges. Notable spectral differences were observed in the 1184–1230 cm$^{-1}$ range, corresponding to carbohydrate structural components like xylan. Huanglongbing -infected leaves displayed decreased band intensities at 1184, 1218, and 1226 cm$^{-1}$ compared to healthy samples. In healthy grapefruit leaves, the 1218 cm$^{-1}$ band was more intense than the 1226 cm$^{-1}$ band, but this ratio was reversed in Huanglongbing -infected leaves. The 1184 cm$^{-1}$ band was more pronounced in nutrient-deficient grapefruit leaves and diminished in nutrient-deficient orange leaves, indicating species-specific responses. Moreover, the 1654 cm$^{-1}$ band—linked to protein content—remained stable in Huanglongbing -infected leaves but significantly increased in nutrient-deficient leaves of both citrus types, further supporting the differentiation of biotic and abiotic stresses [39].

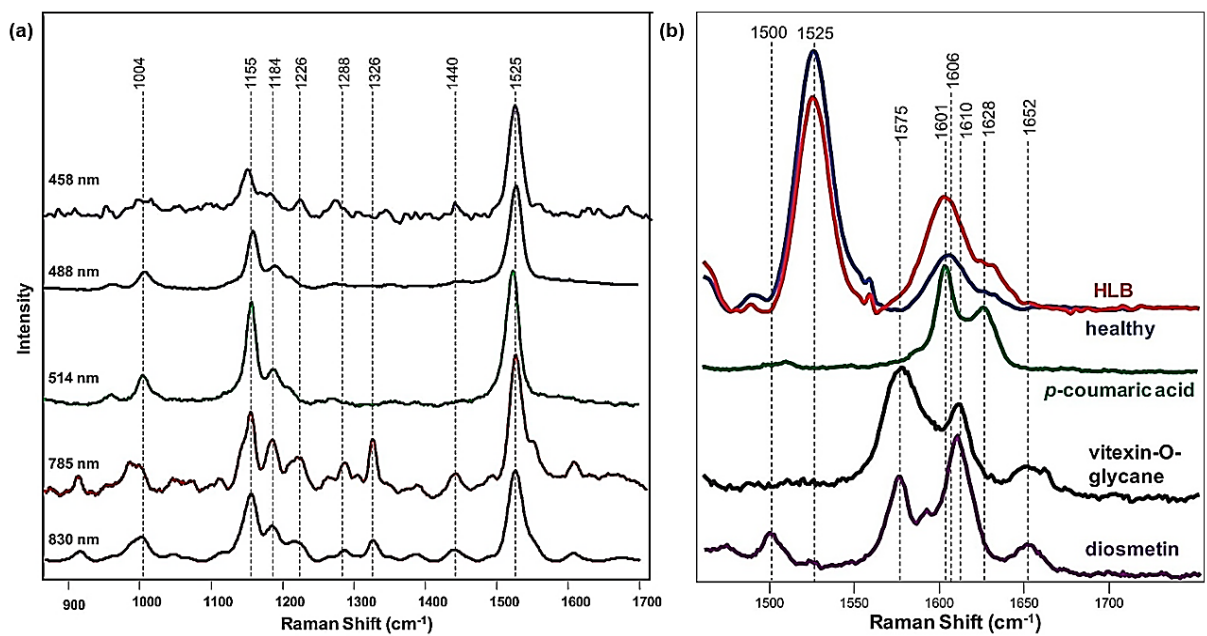

**Figure 6:** Raman spectra of citrus leaves collected at various excitation wavelengths to show laser-dependent spectral variations. Also shown are spectra of healthy and Huanglongbing-infected leaves, along with reference spectra of p-coumaric acid, vitexin-2-O-glucane, and diosmetin [89].

Using a Raman spectrometer equipped with an 831 nm laser, citrus leaf samples were analyzed from healthy, early asymptomatic Huanglongbing, Huanglongbing with blight symptoms, and canker-infected trees. In Huanglongbing-positive orange trees, increased Raman intensities were recorded in the 1601–1630 and 1440–1455 cm$^{-1}$ bands, corresponding to lignin and aliphatic compounds. In contrast, vibrational band intensities at 1184, 1218, and 1226 cm$^{-1}$ were significantly reduced in Huanglongbing-infected leaves compared to other categories, suggesting a decrease in xylan and carbohydrate-related structures. Canker-infected leaves showed reduced lignin band intensities relative to healthy ones, indicating a distinct biochemical profile. Predictive Component (PC) modeling was applied to identify key spectral biomarkers for disease classification. The carotenoid peak at 1525 cm$^{-1}$, lignin peaks at 1601 and 1630 cm$^{-1}$, cellulose bands at 915 and 1326 cm$^{-1}$, xylan at 1184 cm$^{-1}$, and hydrocarbon bands at 1440, 1455, and 1488 cm$^{-1}$ were identified as principal features. The 1630 cm$^{-1}$ peak served as a combined marker for cellulose and lignin. These spectral predictors matched qualitative findings and demonstrated strong potential for distinguishing Huanglongbing and canker infections through spectral fingerprinting [57].

Microscopic confocal Raman spectroscopy with a 785 nm excitation laser revealed detailed biochemical variations in leaf tissues during Huanglongbing progression. Raman bands associated with chlorophyll a were detected at 745, 917, and 1328 cm$^{-1}$ with intensity ratios of 12:2:1, 5:2:1, and 10:2:1. Carotenoid and carotene bands were observed at 1004, 1158, 1390, and 1527 cm$^{-1}$ with progressively declining intensity ratios (7:2:1 to 19:3:1). Cuticular wax was identified at 1050 and 1440 cm$^{-1}$. Glucose and sucrose produced a peak at 853 cm$^{-1}$, showing intensity ratios of 1.8:1.3:1 (healthy:asymptomatic:symptomatic), while starch-related peaks at 866 and 942 cm$^{-1}$ had inverse ratios of 1.3:1.8:1 and 2:5:1. Additional starch bands at 1082, 1250, and 1455 cm$^{-1}$ were absent in healthy but present in asymptomatic and symptomatic leaves. In the 1560–1722 cm$^{-1}$ region, Raman intensity decreased from symptomatic to healthy leaves. Polyphenol-associated peaks at 1577, 1607, and 1630 cm$^{-1}$ showed ratios of 9:8:1, 5:3:2, and 6:4:1, respectively, indicating elevated polyphenol accumulation under Huanglongbing stress. Overall, chlorophyll, carotene, glucose, and sucrose levels declined, while starch and polyphenol contents increased with disease progression, highlighting Raman spectroscopy's capability in tracking subtle biochemical transitions in Huanglongbing -infected leaves [92].

Despite promising advancements, the current body of literature on biochemical-based spectral diagnostics for plant disease remains limited. Many spectral markers and biochemical pathways associated with plant stress, particularly those linked to early Huanglongbing infection, are still poorly understood. Moreover, low-cost, field-deployable Raman spectrometers capable of providing real-time diagnostics are yet to be widely developed or validated. Consequently, further extensive research is required to expand our understanding of spectral biomarkers, improve the resolution of prediction models, and develop robust field-based Raman spectroscopy systems. This review underscores the critical need for scientific innovation in predictive modeling and portable instrumentation, thereby offering valuable direction for future studies aimed at improving early disease detection and sustainable agricultural practices.

**Table 1: Summary of Raman techniques for disease detection in citrus trees.**

| Plants | Disease | Excitation Laser | Emission Range | Statistics | Interpretations |
|---|---|---|---|---|---|
| Sweet Orange, Persian Lime, and Mexican Lime (2016) | Huanglongbing | 785 nm | 800-1800 cm$^{-1}$ | Principal Component Analysis-Linear Discriminant Analysis (PCA-LDA) | PCA–LDA accurately identified Huanglongbing-positive oranges with 86.9% sensitivity, 91.4% specificity, and 89.2% precision [90]. |
| Orange and Grapefruit (2019) | Huanglongbing | 831 nm | 600-1800 cm$^{-1}$ | Orthogonal Partial Least Squares Discriminant Analysis (OPLS-DA) | Raman diagnostics achieved 98% grapefruit, 87% orange accuracy, with 100% and 94% early vs late Huanglongbing detection, respectively [39]. |
| Orange (2019) | Huanglongbing and Citrus Canker | 831 nm | 600-1800 cm$^{-1}$ | Orthogonal Partial Least Squares Discriminant Analysis (OPLS-DA) | Raman spectroscopy with chemometric analysis detects and identifies secondary diseases like blight in Huanglongbing-infected orange trees effectively [57] |
| Orange (2019) | Huanglongbing | 785 nm | 600-1800 cm$^{-1}$ | Partial Least Squares Discrimination Analysis (PLS-DA) and Back-Propagation-Artificial Neural Network (BP-ANN) | PLS-DA showed strong clustering (97.2% accuracy); BP-ANN achieved high prediction performance ($R^2$ = 0.9598, RMSE = 0.0616) [92]. |
| Orange and Grapefruit (2020) | Huanglongbing | 830 nm | 600-1800 cm$^{-1}$ | Orthogonal Partial Least Squares Discriminant Analysis (OPLS-DA) | Raman spectroscopy enabled early Huanglongbing detection, identifying qPCR-negative infected plants via distinct spectral signatures using multivariate analysis [44]. |
| Citrus leaves (2020) | Huanglongbing | 785 nm | 90–3500 cm$^{-1}$ | Partial Least Squares Discrimination Analysis (PLS-DA) | C=C stretching weakens in nutrient-deficient leaves, decreasing with severity but remains stronger than in pathogen-affected Huanglongbing leaves [93]. |
| Grapefruits (2021) | Huanglongbing | 785 nm | 300–2000 cm$^{-1}$ | Kruskal-Wallis one-way analysis | Bacterial activity increases hydroxycinnamates and flavones, decreases lutein, with changes detected using Raman spectroscopy [89]. |

## 5 Limitations and Challenges

While Raman spectroscopy offers a non-destructive and label-free analytical technique with promising potential for plant disease diagnostics, several critical limitations must be addressed to realize its widespread field application [64]. One of the most prominent challenges is the inherently weak nature of Raman scattering. Since Raman signals are typically $10^{-6}$ to $10^{-8}$ times weaker than the incident light, detecting low-abundance biochemical markers associated with early or subtle disease manifestations becomes exceedingly difficult without employing signal enhancement techniques [80]. Surface-enhanced Raman scattering can amplify the signal, but this adds complexity and requires substrate preparation, which may not be practical in on-site agricultural settings. Another major limitation arises from the fluorescence background emitted by various pigments and compounds naturally present in plant tissues [23]. Chlorophylls, carotenoids, and other secondary metabolites often fluoresce upon laser excitation, producing strong background noise that can obscure the already weak Raman signal. This interference significantly complicates spectral analysis and compromises the ability to distinguish disease-specific spectral features. Although techniques such as shifted-excitation Raman difference spectroscopy or time-gated Raman systems can help mitigate fluorescence, these approaches further increase the cost and complexity of the instrumentation [95].

Economic aspect poses a significant barrier to the broader adoption of Raman spectroscopy in agricultural fields. High-resolution benchtop Raman spectrometers, as well as rugged and portable versions suitable for field deployment, remain costly. The investment required for acquiring and maintaining such systems limits their accessibility, especially for small-scale farmers or institutions in resource-constrained regions. Additionally, the miniaturization of spectrometers often compromises spectral resolution or sensitivity, which can further impact diagnostic accuracy in real-world conditions [65], [96].

Raman spectra of biological tissues, including plant leaves and stems, are typically complex and overlapping, reflecting the diverse chemical composition of these matrices [29]. Differentiating between healthy and diseased tissue requires sophisticated multivariate statistical analysis and chemometric techniques such as machine learning algorithms [7]. The effective application of these tools demands both computational resources and expert knowledge, posing a significant challenge for non-specialists aiming to implement Raman spectroscopy for routine field diagnostics [97]. Another intrinsic limitation of Raman spectroscopy is its shallow penetration depth. Raman signals predominantly originate from the surface or near-surface layers of a sample, which restricts the ability to detect internal pathogens or systemic infections that do not manifest at the surface. In contrast, the NIR spectroscopy can penetrate deeper into biological tissues and may be more suitable for subsurface detection. Thus, relying solely on Raman spectroscopy may result in incomplete or inaccurate diagnosis if disease symptoms are internalized or masked by surface features [93].

Biological variability introduces another layer of complexity to disease identification using Raman spectroscopy [72]. Variations due to plant species, cultivar, age, physiological state, nutrient status, and exposure to environmental stressors can all influence the spectral profile. This natural heterogeneity may obscure disease-related spectral changes or generate false positives/negatives, thereby reducing the reliability of diagnostic outcomes. Establishing robust spectral libraries and normalization protocols is essential, but this requires extensive field trials and data standardization efforts [87]. The successful deployment of Raman spectroscopy in field conditions hinges on the expertise of the operators. Accurate spectral acquisition, baseline correction, and interpretation demand not only a strong understanding of vibrational spectroscopy but also a background in plant pathology to correlate spectral features with biological phenomena. The limited availability of trained personnel can hinder the integration of Raman-based tools into routine plant disease surveillance programs [98].

Field-based measurements are inherently prone to environmental interferences. Ambient lighting conditions, especially sunlight, can introduce stray light into the system and reduce the signal-to-noise ratio. Fluctuations in temperature and humidity can also affect instrument calibration and the physicochemical state of plant tissues, potentially distorting the Raman spectra [99]. Shielding mechanisms and adaptive algorithms are required to compensate for these variations, but they add to the operational complexity. Raman spectroscopy holds significant promise for non-invasive, rapid detection of plant diseases, its current limitations—ranging from technical and economic barriers to biological and environmental variability—necessitate further research and technological refinement. Addressing these challenges through innovations will be crucial for transitioning Raman spectroscopy from laboratory research to robust field-ready diagnostic tools [54], [100].

# 6 Conclusion

Raman spectroscopy offers a transformative approach for the early detection and differentiation of Huanglongbing and citrus canker, enabling the identification of pathogen-induced biochemical changes well before visible symptoms appear. Huanglongbing, characterized by its distinctive vibrational signatures, is associated with alterations in carotenoids, chlorophyll, and phenolic compounds, allowing detection during the latent phase of infection, when management interventions are most effective. In citrus canker, Raman spectral markers that reflect cell wall degradation, altered lignin content, and stress-related metabolites provide a rapid and non-invasive means of disease identification. The integration of portable Raman devices with advanced data analysis tools, such as machine learning, has demonstrated high diagnostic accuracy for both diseases under field conditions. Although fluorescence interference, environmental variability, and model generalisation remain challenges, targeted algorithm optimisation and the expansion of pathogen-specific spectral databases are expected to enhance reliability. By enabling precise, on-site diagnosis of Huanglongbing and citrus canker, Raman spectroscopy can support timely control measures, reduce yield losses, and contribute significantly to sustainable citrus production systems.

**Statements and Declarations:**

- **Funding:** The authors declare that no funds, grants, or other support were received during the preparation of this manuscript.

- **Competing Interests:** The authors have no relevant financial or non-financial interests to disclose.

**References**

[1] F. Morán, M. Herrero-Cervera, S. Carvajal-Rojas, and E. Marco-Noales, "Real-time on-site detection of the three 'Candidatus Liberibacter' species associated with HLB disease: a rapid and validated method," *Front. Plant Sci.*, vol. 14, 2023, doi: 10.3389/fpls.2023.1176513.

[2] M. Khalilzadeh, D. J. Aldrich, H. J. Maree, and A. Levy, "Complex interplay: The interactions between citrus tristeza virus and its host," *Virology*, vol. 603, p. 110388, 2025, doi: https://doi.org/10.1016/j.virol.2024.110388.

[3] P. N. Gaikwad *et al.*, "Roles of metabolites in fruit maturation, HLB-defense regulation and crosstalk between phytohormone signalling pathways in citrus," *Plant Growth Regul.*, vol. 105, no. 3, pp. 619–653, 2025, doi: 10.1007/s10725-025-01308-4.

[4] C. Chiyaka, B. H. Singer, S. E. Halbert, J. G. Morris, and A. H. C. van Bruggen, "Modeling huanglongbing transmission within a citrus tree," *Proc. Natl. Acad. Sci.*, vol. 109, no. 30, pp. 12213–12218, Jul. 2012, doi: 10.1073/pnas.1208326109.

[5] J. G. A. Vieira *et al.*, "'Candidatus Liberibacter asiaticus' infection alters the reflectance profile in asymptomatic citrus plants," *Pest Manag. Sci.*, vol. n/a, no. n/a, Nov. 2024, doi: 10.1002/ps.8528.

[6] Y. Hu, N. Lu, K. Bao, S. Liu, R. Li, and G. Huang, "Swords and shields: the war between Candidatus Liberibacter asiaticus and citrus," *Front. Plant Sci.*, vol. Volume 15, 2025, [Online]. Available: https://www.frontiersin.org/journals/plant-science/articles/10.3389/fpls.2024.1518880

[7] S. Bharadwaj, A. Midha, S. Sharma, G. S. Sidhu, and R. Kumar, "Optical screening of citrus leaf diseases using label-free spectroscopic tools: A review," *J. Agric. Food Res.*, vol. 18, p. 101303, 2024, doi: 10.1016/j.jafr.2024.101303.

[8] D. G. Hall, M. L. Richardson, E.-D. Ammar, and S. E. Halbert, "Asian citrus psyllid, iaphorina citri, vector of citrus huanglongbing disease," *Entomol. Exp. Appl.*, vol. 146, no. 2, pp. 207–223, Feb. 2013, doi: https://doi.org/10.1111/eea.12025.

[9] J. C. Barbosa *et al.*, "The leafhopper Agallia albidula is a vector of a phytoplasma associated with 'Huanglongbing'-like symptoms in citrus in Brazil," *Trop. Plant Pathol.*, vol. 50, no. 1, p. 12, 2025, doi: 10.1007/s40858-025-00712-5.

[10] S. A. Naqvi *et al.*, "Citrus Canker—Distribution, Taxonomy, Epidemiology, Disease Cycle, Pathogen Biology, Detection, and Management: A Critical Review and Future Research Agenda," 2022. doi: 10.3390/agronomy12051075.

[11] J. Abdulridha, O. Batuman, and Y. Ampatzidis, "UAV-Based Remote Sensing Technique to Detect Citrus Canker Disease Utilizing Hyperspectral Imaging and Machine Learning," 2019. doi: 10.3390/rs11111373.

[12] D. Thakuria *et al.*, "Citrus Huanglongbing (HLB): Diagnostic and management options," *Physiol. Mol. Plant Pathol.*, vol. 125, p. 102016, 2023, doi: 10.1016/j.pmpp.2023.102016.

[13] R. A. Blaustein, G. L. Lorca, and M. Teplitski, " Challenges for Managing Candidatus Liberibacter spp. (Huanglongbing Disease Pathogen): Current Control Measures and Future Directions ," *Phytopathology®*, vol. 108, no. 4, pp. 424–435, Oct. 2018, doi: 10.1094/phyto-07-17-0260-rvw.

[14] Ş. Kurt *et al.*, "Molecular identification and pathogenicity of Botryosphaeriaceae species associated with citrus wood diseases in the eastern Mediterranean region of Türkiye," *J. Plant Pathol.*, vol. 107, no. 2, pp. 1077–1089, 2025, doi: 10.1007/s42161-025-01877-3.

[15] P. N. Meena *et al.*, "Integrated Pest Management techniques in a Kinnow mandarin (<em>Citrus reticulata</em> Blanco) orchard with an emphasis on yield improvement," *Heliyon*, vol. 11, no. 4, Feb. 2025, doi: 10.1016/j.heliyon.2025.e42574.

[16] J. Li *et al.*, "Developing Citrus Huanglongbing (HLB) Management Strategies Based on the Severity of Symptoms in HLB-Endemic Citrus-Producing Regions," *Phytopathology®*, vol. 109, no. 4, pp. 582–592, Nov. 2018, doi: 10.1094/PHYTO-08-18-0287-R.

[17] X. Li *et al.*, "Advances and perspectives in biological control of postharvest fungal decay in citrus fruit utilizing yeast antagonists," *Int. J. Food Microbiol.*, vol. 432, p. 111093, 2025, doi:

https://doi.org/10.1016/j.ijfoodmicro.2025.111093.

[18]  X. Yu *et al.*, "Detection of fungal disease in citrus fruit based on hyperspectral imaging," *Inf. Process. Agric.*, 2025, doi: https://doi.org/10.1016/j.inpa.2025.02.006.

[19]  S. Hu *et al.*, "Preparation and application of polycaprolactone/β-cyclodextrin/gamma-Decalactone nanofiber composites in citrus postharvest diseases," *Food Chem.*, vol. 463, p. 141476, 2025, doi: https://doi.org/10.1016/j.foodchem.2024.141476.

[20]  A. Aravinthkumar *et al.*, "Comparative efficacy of GRAS chemicals, botanicals and yeast in controlling green mould and fruit nutritional quality enhancement in Kinnow mandarin (Citrus nobilis Lour x Citrus deliciosa Tenora)," *Sci. Hortic. (Amsterdam).*, vol. 339, p. 113869, 2025, doi: https://doi.org/10.1016/j.scienta.2024.113869.

[21]  P. Sharma and P. Abrol, "Multi-component image analysis for citrus disease detection using convolutional neural networks," *Crop Prot.*, vol. 193, p. 107181, 2025, doi: https://doi.org/10.1016/j.cropro.2025.107181.

[22]  A. Mishra, D. Karimi, R. Ehsani, and L. G. Albrigo, "Evaluation of an active optical sensor for detection of Huanglongbing (HLB) disease," *Biosyst. Eng.*, vol. 110, no. 3, pp. 302–309, 2011, doi: 10.1016/j.biosystemseng.2011.09.003.

[23]  A.-K. Mahlein, "Plant Disease Detection by Imaging Sensors – Parallels and Specific Demands for Precision Agriculture and Plant Phenotyping," *Plant Dis.*, vol. 100, no. 2, pp. 241–251, Sep. 2015, doi: 10.1094/PDIS-03-15-0340-FE.

[24]  R. Dong, A. Shiraiwa, and T. Hayashi, "A Simple Diagnostic Method for Citrus Greening Disease With Deep Learning," *Electron. Commun. Japan*, vol. 108, no. 1, p. e12472, Mar. 2025, doi: https://doi.org/10.1002/ecj.12472.

[25]  R. I. Davis, T. G. Gunua, M. F. Kame, D. Tenakanai, and T. K. Ruabete, "Spread of citrus huanglongbing (greening disease) following incursion into Papua New Guinea," *Australas. Plant Pathol.*, vol. 34, no. 4, pp. 517–524, 2005, doi: 10.1071/AP05059.

[26]  K. Yan *et al.*, "Multiple light sources excited fluorescence image-based non-destructive method for citrus Huanglongbing disease detection," *Comput. Electron. Agric.*, vol. 237, p. 110549, 2025, doi:

https://doi.org/10.1016/j.compag.2025.110549.

[27] R. Rai and M. N. Rai, "Emerging Molecular Tools and Breeding Strategies for Plant Bacterial Disease Management BT - Molecular and Biotechnological Tools for Plant Disease Management," J.-T. Chen, M. Khan, A. Parveen, and J. K. Patra, Eds., Singapore: Springer Nature Singapore, 2025, pp. 403–426. doi: 10.1007/978-981-97-7510-1_14.

[28] S. M. Naser, A. Osama, A. Aswar, A. K. Raied, and R. and Alkowni, "Molecular detection and identification of citrus bent leaf viroid (CBLVd) and hop stunt viroid (HSVd) in Palestine," *Arch. Phytopathol. Plant Prot.*, vol. 58, no. 6, pp. 356–364, Apr. 2025, doi: 10.1080/03235408.2025.2488424.

[29] K. R. Clark and P. and Goldberg Oppenheimer, "Shedding light on forest health: noninvasive early foliar pathogen diagnostics via vibrational spectroscopy molecular fingerprinting," *Appl. Spectrosc. Rev.*, pp. 1–19, doi: 10.1080/05704928.2025.2519048.

[30] J. Hasan and S. Bok, "Plasmonic Fluorescence Sensors in Diagnosis of Infectious Diseases," *Biosensors*, vol. 14, no. 3, 2024, doi: 10.3390/bios14030130.

[31] M. Aslam, A. Rani, B. N. Pant, P. Singh, and G. Pandey, "Molecular Diagnostics of Plant Viruses, Viroids, and Phytoplasma: An Updated Overview BT - Molecular and Biotechnological Tools for Plant Disease Management," J.-T. Chen, M. Khan, A. Parveen, and J. K. Patra, Eds., Singapore: Springer Nature Singapore, 2025, pp. 213–233. doi: 10.1007/978-981-97-7510-1_7.

[32] T. H. Hung, S. C. Hung, C. N. Chen, M. H. Hsu, and H. J. Su, "Detection by PCR of Candidatus Liberibacter asiaticus, the bacterium causing citrus huanglongbing in vector psyllids: Application to the study of vector-pathogen relationships," *Plant Pathol.*, vol. 53, no. 1, pp. 96–102, Feb. 2004, doi: 10.1111/j.1365-3059.2004.00948.x.

[33] F. Mokrini *et al.*, "Major pests and diseases threatening the future of Moroccan citrus cultivation and their potential management strategies," *CABI Rev.*, Jan. 2025, doi: 10.1079/cabireviews.2025.0027.

[34] M. Bar-Joseph, "On the Trail of the Longest Plant RNA Virus: Citrus Tristeza Virus," 2025. doi: 10.3390/v17040508.

[35] K. Abd-Elsalam, A. Bahkali, M. Moslem, O. E. Amin, and L. Niessen, "An optimized protocol for DNA extraction from wheat seeds and loop-mediated isothermal amplification (LAMP) to detect Fusarium

graminearum contamination of wheat grain," 2011. doi: 10.3390/ijms12063459.

[36]	M. K. Prasannakumar, P. B. Parivallal, D. Pramesh, H. B. Mahesh, and E. Raj, "LAMP-based foldable microdevice platform for the rapid detection of Magnaporthe oryzae and Sarocladium oryzae in rice seed," *Sci. Rep.*, vol. 11, no. 1, p. 178, 2021, doi: 10.1038/s41598-020-80644-z.

[37]	T. Hussain *et al.*, "Phytophthora spp.'s effect on citrus industry: Current status, challenges, and emerging control strategies," *CABI Rev.*, Jan. 2025, doi: 10.1079/cabireviews.2025.0020.

[38]	Z. Wei, J. Yixue, and H. and Liu, "Optical imaging combined with artificial intelligence in plant disease detection: a comprehensive review," *Spectrosc. Lett.*, pp. 1–25, doi: 10.1080/00387010.2025.2459246.

[39]	L. Sanchez, S. Pant, Z. Xing, K. Mandadi, and D. Kurouski, "Rapid and noninvasive diagnostics of Huanglongbing and nutrient deficits on citrus trees with a handheld Raman spectrometer," *Anal. Bioanal. Chem.*, vol. 411, no. 14, pp. 3125–3133, 2019, doi: 10.1007/s00216-019-01776-4.

[40]	Y. Xia, Y. Xu, J. Li, C. Zhang, and S. Fan, "Recent advances in emerging techniques for non-destructive detection of seed viability: A review," *Artif. Intell. Agric.*, vol. 1, pp. 35–47, 2019, doi: 10.1016/j.aiia.2019.05.001.

[41]	N. Altangerel *et al.*, "In vivo diagnostics of early abiotic plant stress response via Raman spectroscopy," *Proc. Natl. Acad. Sci.*, vol. 114, no. 13, pp. 3393–3396, Mar. 2017, doi: 10.1073/pnas.1701328114.

[42]	A. R. Mishra, D. Karimi, R. Ehsani, and W. S. Lee, "Identification of Citrus Greening (HLB) Using a VIS-NIR Spectroscopy Technique," *Trans. ASABE*, vol. 55, no. 2, pp. 711–720, 2012, doi: 10.13031/2013.41369.

[43]	K. Yan *et al.*, "Three-dimensional fluorescence spectral characteristic of flavonoids for citrus Huanglongbing disease early detection," *Microchem. J.*, vol. 208, p. 112263, 2025, doi: https://doi.org/10.1016/j.microc.2024.112263.

[44]	L. Sanchez, S. Pant, K. Mandadi, and D. Kurouski, "Raman Spectroscopy vs Quantitative Polymerase Chain Reaction In Early Stage Huanglongbing Diagnostics," *Sci. Rep.*, vol. 10, no. 1, p. 10101, 2020, doi: 10.1038/s41598-020-67148-6.

[45]	P. L. Vishnuppriyan, R. Karthik, B. A. Prabu, and K. Agarwal, "A dual path deep-learning network with multi-scale cross attention and pyramid vision transformer for citrus leaf disease detection," *Eur. Phys. J.*

*Spec. Top.*, 2025, doi: 10.1140/epjs/s11734-025-01731-8.

[46] R. Ramesh kumar *et al.*, "Assessment of huanglongbing induced changes in primary and secondary metabolites of acid lime," *Physiol. Mol. Plant Pathol.*, vol. 136, p. 102547, 2025, doi: https://doi.org/10.1016/j.pmpp.2024.102547.

[47] K. Wu, E. D. Vu, S. Ghosh, R. Mishra, and B. C. Bonning, "Continuous cell lines derived from the Asian citrus psyllid, Diaphorina citri, harbor viruses and Wolbachia," *Sci. Rep.*, vol. 15, no. 1, p. 124, 2025, doi: 10.1038/s41598-024-83671-2.

[48] S. Yan, Z. Huang, S. Gao, and Y. Kang, "Dynamics of a citrus disease model with intercropping and insecticides: application to Huanglongbing," *Adv. Contin. Discret. Model.*, vol. 2025, no. 1, p. 98, 2025, doi: 10.1186/s13662-025-03962-4.

[49] E. Etxeberria, P. Gonzalez, D. Achor, and G. Albrigo, "Anatomical distribution of abnormally high levels of starch in HLB-affected Valencia orange trees," *Physiol. Mol. Plant Pathol.*, vol. 74, no. 1, pp. 76–83, 2009, doi: https://doi.org/10.1016/j.pmpp.2009.09.004.

[50] N. Esiobu *et al.*, "Rhizosphere Microbiomes of Citrus Plants in Historically Undisturbed 100-Year-Old Groves Appear to Mitigate Susceptibility to Citrus Greening Disease," 2025. doi: 10.3390/microorganisms13040763.

[51] B. Hu *et al.*, "Molecular signatures between citrus and Candidatus Liberibacter asiaticus," *PLOS Pathog.*, vol. 17, no. 12, p. e1010071, Dec. 2021, [Online]. Available: https://doi.org/10.1371/journal.ppat.1010071

[52] M. Wang, Z. Li, and J. Zhao, "Citrus Greening Disease Infection Reduces the Energy Flow Through Soil Nematode Food Webs," 2025. doi: 10.3390/agronomy15030635.

[53] S. Sankaran, R. Ehsani, and E. Etxeberria, "Mid-infrared spectroscopy for detection of Huanglongbing (greening) in citrus leaves," *Talanta*, vol. 83, no. 2, pp. 574–581, 2010, doi: 10.1016/j.talanta.2010.10.008.

[54] W. S. Lee and R. Ehsani, "Sensing systems for precision agriculture in Florida," *Comput. Electron. Agric.*, vol. 112, pp. 2–9, 2015, doi: 10.1016/j.compag.2014.11.005.

[55] G. McCollum and E. Baldwin, "Huanglongbing: Devastating Disease of Citrus," in *Horticultural*

*Reviews, Volume 44*, 2016, pp. 315–361. doi: https://doi.org/10.1002/9781119281269.ch7.

[56] S. Sankaran, A. Mishra, R. Ehsani, and C. Davis, "A review of advanced techniques for detecting plant diseases," *Comput. Electron. Agric.*, vol. 72, no. 1, pp. 1–13, 2010, doi: https://doi.org/10.1016/j.compag.2010.02.007.

[57] L. Sanchez, S. Pant, M. Irey, K. Mandadi, and D. Kurouski, "Detection and identification of canker and blight on orange trees using a hand-held Raman spectrometer," *J. Raman Spectrosc.*, vol. 50, no. 12, pp. 1875–1880, Dec. 2019, doi: https://doi.org/10.1002/jrs.5741.

[58] I. Pence and A. Mahadevan-Jansen, "Clinical instrumentation and applications of Raman spectroscopy," *Chem. Soc. Rev.*, vol. 45, no. 7, pp. 1958–1979, 2016, doi: 10.1039/C5CS00581G.

[59] R. S. Das and Y. K. Agrawal, "Raman spectroscopy: Recent advancements, techniques and applications," *Vib. Spectrosc.*, vol. 57, no. 2, pp. 163–176, 2011, doi: https://doi.org/10.1016/j.vibspec.2011.08.003.

[60] X. Zhu, X. Tao, L. Qingyu, and Y. and Duan, "Technical Development of Raman Spectroscopy: From Instrumental to Advanced Combined Technologies," *Appl. Spectrosc. Rev.*, vol. 49, no. 1, pp. 64–82, Jan. 2014, doi: 10.1080/05704928.2013.798801.

[61] J. Coates, "Vibrational Spectroscopy: Instrumentation for Infrared and Raman Spectroscopy∗," *Appl. Spectrosc. Rev.*, vol. 33, no. 4, pp. 267–425, Nov. 1998, doi: 10.1080/05704929808002060.

[62] A. Saletnik, B. Saletnik, and C. Puchalski, "Overview of Popular Techniques of Raman Spectroscopy and Their Potential in the Study of Plant Tissues," 2021. doi: 10.3390/molecules26061537.

[63] W. Z. Payne and D. Kurouski, "Raman spectroscopy enables phenotyping and assessment of nutrition values of plants: a review," *Plant Methods*, vol. 17, no. 1, p. 78, 2021, doi: 10.1186/s13007-021-00781-y.

[64] N. Gierlinger and M. Schwanninger, "The potential of Raman microscopy and Raman imaging in plant research," *J. Spectrosc.*, vol. 21, no. 2, p. 498206, Jan. 2007, doi: https://doi.org/10.1155/2007/498206.

[65] A. Saletnik, B. Saletnik, G. Zaguła, and C. Puchalski, "Raman Spectroscopy for Plant Disease Detection in Next-Generation Agriculture," 2024. doi: 10.3390/su16135474.

[66] C. Orecchio, C. Sacco Botto, E. Alladio, C. D'Errico, M. Vincenti, and E. Noris, "Non-invasive and early detection of tomato spotted wilt virus infection in tomato plants using a hand-held Raman spectrometer

and machine learning modelling," *Plant Stress*, vol. 15, p. 100732, 2025, doi: https://doi.org/10.1016/j.stress.2024.100732.

[67] S. Jeong and H. Chung, "Combining two-trace two-dimensional correlation analysis and convolutional autoencoder-based feature extraction from an entire correlation map to enhance vibrational spectroscopic discrimination of geographical origins of agricultural products," *Talanta*, vol. 285, p. 127385, 2025, doi: https://doi.org/10.1016/j.talanta.2024.127385.

[68] M. Wang, C. Zhang, S. Yan, T. Chen, H. Fang, and X. Yuan, "Wide-Field Super-Resolved Raman Imaging of Carbon Materials," *ACS Photonics*, vol. 8, no. 6, pp. 1801–1809, Jun. 2021, doi: 10.1021/acsphotonics.1c00392.

[69] K. Müller *et al.*, "Raman Microspectroscopy to Trace the Incorporation of Deuterium from Labeled (Micro)Plastics into Microbial Cells," *Anal. Chem.*, vol. 97, no. 8, pp. 4440–4451, Mar. 2025, doi: 10.1021/acs.analchem.4c05827.

[70] N. Gierlinger, T. Keplinger, and M. Harrington, "Imaging of plant cell walls by confocal Raman microscopy," *Nat. Protoc.*, vol. 7, no. 9, pp. 1694–1708, 2012, doi: 10.1038/nprot.2012.092.

[71] J. Guo *et al.*, "Exploring the potential of microscopic hyperspectral, Raman, and LIBS for nondestructive quality assessment of diverse rice samples," *Plant Methods*, vol. 21, no. 1, p. 25, 2025, doi: 10.1186/s13007-025-01345-0.

[72] B. P. Mateu, P. Bock, and N. Gierlinger, "Raman Imaging of Plant Cell Walls BT  - The Plant Cell Wall: Methods and Protocols," Z. A. Popper, Ed., New York, NY: Springer New York, 2020, pp. 251–295. doi: 10.1007/978-1-0716-0621-6_15.

[73] H. Zhang, L. Wang, W. Li, L. Shao, and J. Hu, "Optical Spectroscopy for Sustainable Agriculture: Crop and Soil Management BT  - Agriculture Value Chain — Challenges and Trends in Academia and Industry: RUC-APS Volume 2," J. Hernández and J. Kacprzyk, Eds., Cham: Springer Nature Switzerland, 2025, pp. 171–188. doi: 10.1007/978-3-031-70745-2_12.

[74] Z. Yang, S. Lv, S. Zeng, S. Xia, and H. Li, "A novel reconstruction method for maize component visualization in low-quality Raman hyperspectral images," *Microchem. J.*, vol. 210, p. 113049, 2025, doi: https://doi.org/10.1016/j.microc.2025.113049.

[75] S. H. Lee, N. C. Lindquist, N. J. Wittenberg, L. R. Jordan, and S. H. Oh, "Real-time full-spectral imaging and affinity measurements from 50 microfluidic channels using nanohole surface plasmon resonance.," *Lab Chip*, vol. 12, no. 20, pp. 3882–3890, 2012, doi: 10.1039/c2lc40455a.

[76] M. Kizilov, V. Cheburkanov, J. Harrington, and V. Yakovlev, "Advanced preprocessing and analysis techniques for enhanced Raman spectroscopy data interpretation," in *Proc.SPIE*, Mar. 2025, p. 133110F. doi: 10.1117/12.3048830.

[77] Chad A Lieber and Anita Mahadevan-Jansen, "Automated Method for Subtraction of Fluorescence from Biological Raman Spectra," *Appl. Spectrosc.*, vol. 57, no. 11, pp. 1363–1367, Nov. 2003, doi: 10.1366/000370203322554518.

[78] J. Chen, J. M. Reyes, R. Schiemer, G. Wang, J. Studts, and M. Franzreb, "Digital Butterworth filter as preprocessing method for implementing Raman spectroscopy as an analytical method in downstream processing of biopharmaceuticals," *J. Chromatogr. A*, vol. 1756, p. 466069, 2025, doi: https://doi.org/10.1016/j.chroma.2025.466069.

[79] Jianhua Zhao, Harvey Lui, David I McLean, and Haishan Zeng, "Automated Autofluorescence Background Subtraction Algorithm for Biomedical Raman Spectroscopy," *Appl. Spectrosc.*, vol. 61, no. 11, pp. 1225–1232, Nov. 2007, doi: 10.1366/000370207782597003.

[80] Y. Liu *et al.*, "Artificial intelligence guided Raman spectroscopy in biomedicine: Applications and prospects," *J. Pharm. Anal.*, p. 101271, 2025, doi: https://doi.org/10.1016/j.jpha.2025.101271.

[81] A. Savitzky and M. J. E. Golay, "Smoothing and Differentiation of Data by Simplified Least Squares Procedures.," *Anal. Chem.*, vol. 36, no. 8, pp. 1627–1639, Jul. 1964, doi: 10.1021/ac60214a047.

[82] D. Boateng, "Advances in deep learning-based applications for Raman spectroscopy analysis: A mini-review of the progress and challenges," *Microchem. J.*, vol. 209, p. 112692, 2025, doi: https://doi.org/10.1016/j.microc.2025.112692.

[83] Å. Rinnan, F. van den Berg, and S. B. Engelsen, "Review of the most common pre-processing techniques for near-infrared spectra," *TrAC Trends Anal. Chem.*, vol. 28, no. 10, pp. 1201–1222, 2009, doi: https://doi.org/10.1016/j.trac.2009.07.007.

[84] N. K. Afseth and A. Kohler, "Extended multiplicative signal correction in vibrational spectroscopy, a

tutorial," *Chemom. Intell. Lab. Syst.*, vol. 117, pp. 92–99, 2012, doi: https://doi.org/10.1016/j.chemolab.2012.03.004.

[85] D. Cozzolino, P. J. Williams, and L. C. Hoffman, "An overview of pre-processing methods available for hyperspectral imaging applications," *Microchem. J.*, vol. 193, p. 109129, 2023, doi: https://doi.org/10.1016/j.microc.2023.109129.

[86] P. Bassan et al., "Resonant Mie Scattering (RMieS) correction of infrared spectra from highly scattering biological samples," *Analyst*, vol. 135, no. 2, pp. 268–277, 2010, doi: 10.1039/B921056C.

[87] S. Fang, R. Cui, Y. Wang, Y. Zhao, K. Yu, and A. Jiang, "Application of multiple spectral systems for the tree disease detection: A review," *Appl. Spectrosc. Rev.*, vol. 58, no. 2, pp. 83–109, Feb. 2023, doi: 10.1080/05704928.2021.1930552.

[88] M. S. Hossain et al., "Automatic Navigation and Self-Driving Technology in Agricultural Machinery: A State-of-the-Art Systematic Review," *IEEE Access*, vol. 13, pp. 94370–94401, 2025, doi: 10.1109/ACCESS.2025.3573324.

[89] T. Dou et al., "Biochemical Origin of Raman-Based Diagnostics of Huanglongbing in Grapefruit Trees," *Front. Plant Sci.*, vol. Volume 12, 2021, [Online]. Available: https://www.frontiersin.org/journals/plant-science/articles/10.3389/fpls.2021.680991

[90] Moisés Roberto Vallejo Pérez, María Guadalupe Galindo Mendoza, Miguel Ghebre Ramírez Elías, Francisco Javier González, Hugo Ricardo Navarro Contreras, and Carlos Contreras Servín, "Raman Spectroscopy an Option for the Early Detection of Citrus Huanglongbing," *Appl. Spectrosc.*, vol. 70, no. 5, pp. 829–839, Mar. 2016, doi: 10.1177/0003702816638229.

[91] M. T. Ahmed, M. W. Ahmed, and M. Kamruzzaman, "A systematic review of explainable artificial intelligence for spectroscopic agricultural quality assessment," *Comput. Electron. Agric.*, vol. 235, p. 110354, 2025, doi: https://doi.org/10.1016/j.compag.2025.110354.

[92] K. Wang, Y. Liao, Y. Meng, X. Jiao, W. Huang, and T. C. Liu, "The Early, Rapid, and Non-Destructive Detection of Citrus Huanglongbing (HLB) Based on Microscopic Confocal Raman," *Food Anal. Methods*, vol. 12, no. 11, pp. 2500–2508, 2019, doi: 10.1007/s12161-019-01598-1.

[93] Y. Liu et al., "Diagnosis of Citrus Greening using Raman Spectroscopy-Based Pattern Recognition," *J.*

*Appl. Spectrosc.*, vol. 87, no. 1, pp. 150–158, 2020, doi: 10.1007/s10812-020-00976-6.

[94] J. Cai, C. Zou, L. Yin, S. Jiang, H. R. El-Seedi, and Z. Guo, "Characterization and recognition of citrus fruit spoilage fungi using Raman scattering spectroscopic imaging," *Vib. Spectrosc.*, vol. 124, p. 103474, 2023, doi: https://doi.org/10.1016/j.vibspec.2022.103474.

[95] C. Farber and D. Kurouski, "Detection and Identification of Plant Pathogens on Maize Kernels with a Hand-Held Raman Spectrometer," *Anal. Chem.*, vol. 90, no. 5, pp. 3009–3012, Mar. 2018, doi: 10.1021/acs.analchem.8b00222.

[96] S. Weng *et al.*, "Advanced Application of Raman Spectroscopy and Surface-Enhanced Raman Spectroscopy in Plant Disease Diagnostics: A Review," *J. Agric. Food Chem.*, vol. 69, no. 10, pp. 2950–2964, Mar. 2021, doi: 10.1021/acs.jafc.0c07205.

[97] K. B. Bec, J. Grabska, G. K. Bonn, M. Popp, and C. W. Huck, "Principles and Applications of Vibrational Spectroscopic Imaging in Plant Science: A Review," *Front. Plant Sci.*, vol. Volume 11, 2020, [Online]. Available: https://www.frontiersin.org/journals/plant-science/articles/10.3389/fpls.2020.01226

[98] H. Schulz and M. Baranska, "Identification and quantification of valuable plant substances by IR and Raman spectroscopy," *Vib. Spectrosc.*, vol. 43, no. 1, pp. 13–25, 2007, doi: https://doi.org/10.1016/j.vibspec.2006.06.001.

[99] A. Issatayeva *et al.*, "SERS-based methods for the detection of genomic biomarkers of cancer," *Talanta*, vol. 267, no. September 2023, p. 125198, 2024, doi: 10.1016/j.talanta.2023.125198.

[100] W. Z. Payne and D. Kurouski, "Raman-Based Diagnostics of Biotic and Abiotic Stresses in Plants. A Review," *Front. Plant Sci.*, vol. Volume 11, 2021, [Online]. Available: https://www.frontiersin.org/journals/plant-science/articles/10.3389/fpls.2020.616672